\documentclass[twocolumn,preprintnumbers,amsmath,amssymb,amsfonts,prb,floatfix]{revtex4}

\usepackage{graphicx} 
\usepackage{dcolumn} 
\usepackage{bm} 

\begin{document}

\title{Study of the topological Hall effect on simple models}  

\author{G.~Metalidis}
  \email{georgo@mpi-halle.de}
\author{P.~Bruno}
  \email{bruno@mpi-halle.de}
\affiliation{%
Max-Planck-Institut f\"{u}r Mikrostrukturphysik, Weinberg 2,
D-06120 Halle, Germany} \homepage{http://www.mpi-halle.de}
\date{\today}

\begin{abstract}
Recently, a chirality-driven contribution to the anomalous Hall
effect has been found that is induced by the Berry phase and does
not directly involve spin-orbit coupling. In this paper, we will
investigate this effect numerically in a two-dimensional electron
gas with a simple magnetic texture model. Both the adiabatic and
non-adiabatic regimes are studied, including the effect of
disorder. By studying the transition between both regimes the
discussion about the correct adiabaticity criterium in the
diffusive limit is clarified.
\end{abstract}

\maketitle

\section{Introduction}
When studying the Hall effect in ferromagnets, an anomalous
contribution was found giving rise to a non-zero Hall effect even
in the absence of an externally applied magnetic field. Spin-orbit
coupling was invoked to explain this effect, as it gives rise to
two scattering mechanisms (skew scattering~\cite{Karplus, Smit}
and side jump~\cite{Berger}) leading to a preferential scattering
direction that is different for spin-up and spin-down. Since the
spin subbands in ferromagnets are unequally populated, this
spin-dependent scattering can indeed give rise to a charge Hall
effect in ferromagnets, while in normal semiconductors these
mechanisms explain the much-discussed spin Hall
effect~\cite{Hirsch}.

Quite recently, one has found that an anomalous Hall effect can
exist even in the absence of spin-orbit interaction in some
classes of frustrated ferromagnets like pyrochlore-type compounds
with non-coplanar magnetic moments in the elementary
cell~\cite{Taguchi}. In these systems, the Hall effect is a result
of the Berry phase~\cite{Berry} an electron acquires when moving
in a system with nontrivial (chiral) spin texture. Since the
effect can be attributed to the topology of the magnetization
field, the term topological Hall effect was coined~\cite{Bruno}.
Theoretical calculations on this effect~\cite{Ohgushi, Nagaosa}
have mainly concentrated on the adiabatic regime, where the
electron spin aligns perfectly with the local magnetization during
its movement. In this regime, by doing a transformation that
aligns the quantization axis with the local magnetization
direction at every point in space, the original Hamiltonian can be
mapped onto a model of spinless electrons moving in an effective
vector potential~\cite{Ohgushi, Bruno}. However, only a few papers
have dealt with the non-adiabatic limit (using perturbation
theory~\cite{Tatara}) and even less is known about the transition
between the two regimes.

In this paper, we will show numerical calculations on the
topological Hall effect in a two-dimensional electron gas (2DEG).
It was proposed that in this system an artificial chirality can be
created e.g. by the stray field of a lattice of ferromagnetic
nanocylinders placed above the 2DEG, having the advantage that all
relevant parameters are controllable up to some
extent~\cite{Bruno}. In the current paper however, we will only
consider some very simple model chiralities in the system.
Nevertheless, with these models we are able to investigate
numerically the existence of the topological Hall effect. More
importantly, the transition point between the non-adiabatic and
adiabatic regime will be studied for different values of the
elastic mean free path. In the diffusive regime, it is found that
the transition occurs slower as we increase the amount of
disorder. This sheds some light on a long-standing discussion
about the relevant adiabaticity criterium; it is in agreement with
the one derived by Stern~\cite{Stern} and speaks against the
criterium put forward by Loss~\cite{Loss}.

The paper is subdivided as follows. In the next section, we will
first give a more detailed description of the topological Hall
effect, discussing the transformation of the original Hamiltonian
in the adiabatic limit onto a model of spinless electrons moving
in an effective vector potential. After different adiabaticity
criteria are presented in section~\ref{adiabaticity}, a discussion
concerning the calculation of Hall resistances and resistivities
follows. In particular, a special phase averaging method is
introduced for describing large (i.e. system length $L$ $\gg$
phase coherence length $L_\phi$) disordered systems; technical
details of this method are given in the appendix. The main results
of the paper can be found in section~\ref{Results}.

\section{Topological Hall Effect} \label{topologicalHall}
In order to explain what the topological Hall effect is about, we
will consider a two-dimensional electron gas (2DEG), subjected to
a spatially varying magnetization $\mathbf{M}(\mathbf{r})$. The
Hamiltonian of such a system can be written as:
\begin{equation} \label{Hamiltonian1}
H =  - \frac{\hbar^2}{2 m^\ast} \nabla^2 - g \, \mathbf{\sigma}
\cdot \mathbf{M}(\mathbf{r}),
\end{equation}
with $m^\ast$ the effective electron mass, $\mathbf{\sigma}$ the
vector of Pauli spin matrices, and $g$ a coupling constant. We
assume that the amplitude of the magnetization is constant; only
its direction is position dependent: $\mathbf{M}(\mathbf{r}) = M
\mathbf{n}(\mathbf{r})$, with $\mathbf{n}$ a vector of unit
amplitude.

For our numerical purposes, it is necessary to obtain a
tight-binding description of this system by discretizing the
Schr\"{o}dinger equation on a square lattice with lattice constant
$a$. The tight-binding equivalent of Eq.~(\ref{Hamiltonian1})
reads
\begin{equation}
H = - t \sum_{\langle i j \rangle} \sum_{\alpha} | i \alpha
\rangle \langle j \alpha | - g M \sum_{i} \sum_{\alpha, \beta} | i
\alpha \rangle \, \mathbf{\sigma}_{\alpha \beta} \cdot
\mathbf{n}_i \, \langle i \beta |,
\end{equation}
where $i,j$ label the lattice sites, $\alpha, \beta$ are spin
indices, $t = \frac{\hbar^2}{2 m a^2}$ is the hopping amplitude,
and $\mathbf{n}_i = \mathbf{n}(\mathbf{r}_i)$. The first summation
runs over nearest neighbors.

\begin{figure}
\includegraphics[width = 7.5 cm]{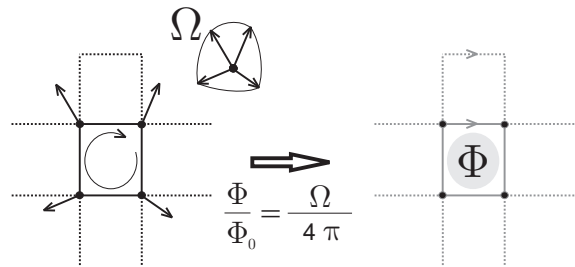}
\caption{ \label{Fig1} Mapping of the tight-binding Hamiltonian
for an electron in a magnetic texture to a spinless electron
moving around a fluxtube. Grey color of the vertices on the right
picture stand for the change in hopping amplitude, while the
arrows denote a change in hopping phase due to the fluxtube.}
\end{figure}

When the coupling $g$ is large enough, the adiabatic limit is
reached meaning that spin flip transitions will be absent and the
electron spin will stay aligned to the local magnetization
direction. In that case, the solutions for the spin-up spinors
(with respect to the local magnetization) are given by
\begin{equation}
| \chi_i \rangle = \left (
\begin{array}{l}
\cos \left( \frac{\theta_i}{2} \right) \exp \left( -\mathrm{i}
\frac{\phi_i}{2} \right) \\
\sin \left( \frac{\theta_i}{2} \right) \exp \left( \mathrm{i}
\frac{\phi_i}{2} \right),
\end{array}
\right )
\end{equation}
with $(\theta_i, \phi_i)$ the spherical coordinates of the local
magnetization direction $\mathbf{n}_i$. By projecting the
Hamiltonian onto the subspace spanned by these spinors, one can
derive an effective Hamiltonian~\cite{Anderson, Ohgushi}
\begin{equation} \label{Heff}
H_\text{eff} = \sum_{\langle i,j \rangle} t^\text{eff}_{ij} | i
\rangle \langle j |,
\end{equation}
where the hopping amplitudes are now
\begin{equation} \label{teff}
t^\text{eff}_{ij} = -t \cos \left( \frac{\zeta_{ij}}{2} \right)
\exp \left( \mathrm{i} \gamma_{ij} \right).
\end{equation}
In this expresssion, $\zeta_{ij}$ is the angle between the vectors
$\mathbf{n}_i$ and $\mathbf{n}_j$. The angle $\gamma_{ij}$ needs
some more explanation. Suppose an electron makes a closed
trajectory around a lattice cell like depicted in Fig.~\ref{Fig1}.
Since its spin follows adiabatically the local magnetization
direction, the electron will pick up a Berry phase that is equal
to half the solid angle $\Omega$ subtended by the magnetization
directions at the four corners of the cell~\cite{Berry}, and the
quantity $\gamma_{ij}$ is related to this Berry phase.

To make things more clear, let's consider a flux
\begin{equation} \label{topflux}
\Phi / \Phi_0 = \frac{\Omega}{4 \pi}
\end{equation}
piercing through a lattice cell and forget about the magnetization
for a moment ($\Phi_0 = h/e$ is the magnetic flux quantum). The
influence of such a fluxtube can be described in a tight-binding
model by a Peierls substitution,~\cite{Peierls} which would change
the hopping phase as $t \rightarrow t \exp(- \mathrm{i} e/ \hbar
\int \mathbf{A} \cdot \mathrm{d}\mathbf{l})$, with the integral
evaluated along the hopping path. This is similar to the phase
factor in Eq.~(\ref{teff}) with
\begin{equation}
\gamma_{ij} = -e / \hbar \int_i^j \mathbf{A} \cdot
\mathrm{d}\mathbf{l},
\end{equation}
meaning that in principle the Berry phase effect can be described
by an effective vector potential. E.g., one could make a choice of
gauge for $\mathbf{A}$ such that the hopping phase on all the
vertices above the fluxtube change as $t \rightarrow t \exp \left(
\mathrm{i} 2 \pi \Phi / \Phi_0 \right) = t \exp \left( \mathrm{i}
\Omega / 2 \right)$ (see Fig.~\ref{Fig1}). This gauge choice would
correspond to the Landau gauge in a situation where the effective
fluxes through all lattice cells would be equal (i.e. there is a
homogeneous field). In short, the effect of the $\gamma_{ij}$ can
thus be described by having a distribution of effective magnetic
fluxes piercing through the lattice cells. As such, for a given
gauge, the quantities $\gamma_{ij}$ are defined in a unique way.

In summary, the effect of the transformation on the tight-binding
Hamiltonian will be twofold: first, the hopping amplitude between
neighbors changes, depending on the angle between the
magnetization directions, and secondly, there is a Berry phase
effect that effectively can be described in terms of a fluxtube
distribution, where the value of the flux (in units $\Phi_0$)
through a single lattice cell is given by $\frac{\Omega}{4 \pi}$,
and $\Omega$ is the solid angle subtended by the magnetization
directions at the four corners of the lattice cell.

By means of this mapping it is clear now that the magnetization
texture can indeed give rise to a Hall effect. It should be noted
that the effective flux is given by a solid angle, so that the
topological Hall effect can only be nonzero for non-coplanar
textures. Furthermore, there is no need to invoke any kind of
spin-orbit coupling for the topological Hall effect to appear.

\section{Adiabaticity Criterium} \label{adiabaticity}
In order to be able to do the transformation between the
Hamiltonian describing an electron moving in a magnetization
texture [Eq.~(\ref{Hamiltonian1}), which will be referred to as
the magnetization model from now on], and the effective
Hamiltonian in Eq.~(\ref{Heff}) [called the flux model from now
on], the electron spin has to stay aligned to the local
magnetization direction during its movement. This adiabatic regime
will be reached when the spin precession frequency $\omega_s$ is
large compared to the inverse of a timescale $\tau$ that is
characteristic for the rate at which the electron sees the
magnetization direction change.

If transport through the nanostructure is ballistic (i.e. there is
no disorder), then this timescale is given by $\tau = \xi / v_F$,
where $v_F$ is the Fermi velocity of the electrons, and $\xi$ is
the length scale over which the magnetization changes its
direction substantially. In this case, the adiabaticity criterium
is thus
\begin{equation} \label{critball}
Q = \frac{\omega_s \xi}{v_F} \gg 1.
\end{equation}
When introducing disorder in the system, it is clear that this
criterium is still valid as long as the mean free path $l_m$ is
larger than $\xi$. However, when going to the strongly diffusive
regime $l_m < \xi$, two different timescales $\tau$ appear in the
literature and there is still a discussion going on about the
relevant one.~\cite{Stern, Loss, Popp, vanLangen} In analytical
calculations on rings, Stern has found that the relevant timescale
would be the elastic scattering time $\tau = \tau_m$, leading to
what is referred to as the pessimistic criterium in the
literature~\cite{Stern}
\begin{equation} \label{critpess}
Q \gg \frac{\xi}{l_m}.
\end{equation}
However, other papers by Loss and coworkers~\cite{Loss} have put
forward that the relevant timescale is the Thouless time
$\tau_\text{th} = (\xi / l_m)^2 \, \tau_m$, i.e. the time the
particle needs to diffusive through a distance $\xi$. This would
lead to a less stringent adiabaticity criterium since
$\tau_\text{th} > \tau_m$ in the diffusive regime, and it is
therefore known as the optimistic criterium:
\begin{equation}
Q \gg \frac{l_m}{\xi}.
\end{equation}
In a paper of van Langen \textit{et al.} the pessimistic criterium
was confirmed by a semi-classical analysis~\cite{vanLangen}, and
later numerically by Popp \textit{et al.}~\cite{Popp} In
section~\ref{Results}, we will have a closer look at the
transition point between the adiabatic and non-adiabatic regimes
ourselves and find also confirmation for the criterium in
Eq.~(\ref{critpess}).

\section{Calculation of the Hall resistance} \label{Calculation}
\begin{figure}
\includegraphics[width = 5cm]{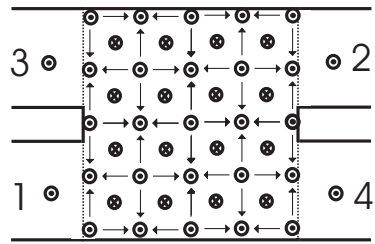}
\caption{ \label{Fig2} Geometry of a single section used to
calculate the Hall resistance. A possible magnetization texture is
shown. In the leads, the magnetization is chosen to point out of
the plane of the paper.}
\end{figure}

In order to observe the topological Hall effect, we will
numerically calculate the Hall resistance in a geometry like shown
in Fig.~\ref{Fig2}. The Hall resistance $R_H$ in this geometry is
defined as
\begin{equation} \label{RH}
R_H = \frac{1}{2} (R_{12,34} - R_{34,12}),
\end{equation}
where we use the common notation
\begin{equation} \label{RHh}
R_{ij, kl} = \frac{V_k - V_l}{I_i}
\end{equation}
for a measurement where current is supplied through contacts $i$
and $j$, and the voltage difference $V_k - V_l$ is measured,
fixing $I_k = I_l = 0$. Making the difference between these two
four-point resistances is equivalent to defining the Hall
resistivity $\rho_H = \frac{1}{2} (\rho_{xy} - \rho_{yx})$ as the
part of the resistivity tensor that is anti-symmetric with respect
to time reversal~\cite{Buettiker}.

Within the Landauer-B\"{u}ttiker (LB) formalism, the four-point
resistances in Eq.~(\ref{RHh}) can be expressed
as~\cite{Buettiker}
\begin{equation}
R_{ij,kl} = \frac{h}{e^2} \frac{T_{ki} T_{lj} - T_{li} T_{kj}}{D}
\end{equation}
as a function of transmission probabilities $T_{pq}$ between the
leads. In this expression, $D$ is an arbitrary $3 \times 3$
subdeterminant of the matrix $A$ relating the currents and the
voltages on the leads: $I_p = \sum_q A_{pq} V_q$.

In principle, one could proceed now by calculating $R_H$ as a
function of the adiabaticity parameter $Q$ and compare the rate of
convergence for samples with different mean free paths in order to
find the relevant adiabaticity criterium in the diffusive regime.
However, since we make use of the LB-formalism, it is assumed that
the whole structure is phase coherent. It is well known that the
resistances in such structures are quantitatively strongly
dependent on the actual configuration of the impurities in the
system even when they are characterized by the same mean free
path. Although this is an integral part of the physics of
mesoscopic systems, it makes a quantitative comparison of $R_H$
between samples with different mean free paths useless. We would
like to compare properties of a macroscopic system, i.e. a system
with a finite phase coherence length in which such fluctuations
are absent.

Therefore, some kind of (phase) averaging over different disorder
configurations has to be introduced to find a description of the
transport properties in terms of a macroscopic material constant,
like the Hall resistivity. Some care should be taken in defining
such an averaging procedure; e.g., just calculating the
mathematical average $R_H = \frac{1}{N}\sum_{i=1}^N R_H^i$ of the
Hall resistances $R_H^i$ found for $N$ different impurity
configurations does not give a quantity that is directly related
to the Hall resistivity $\rho_H$ of a macroscopic system.

The basic idea behind the averaging procedure we have chosen is
that a macroscopic system ($L \gg L_\phi$) can be thought to
consist of smaller phase coherent sections of size $L \approx
L_\phi$. For every smaller section, we can use the LB-formula to
derive its transport properties, and the properties of a
macroscopic system can then be found by attaching such sections in
an incoherent way. In our case, the smaller sections from which a
macroscopic system will be built up are structures with the
geometry in Fig.~\ref{Fig2}. Since a more detailed discussion of
our particular averaging procedure and the corresponding Hall
resistivity $\rho_H$ is rather technical, it is given in a
separate appendix at the end of the paper.

\section{Results} \label{Results}
\begin{figure}
\includegraphics[width = 7.5cm]{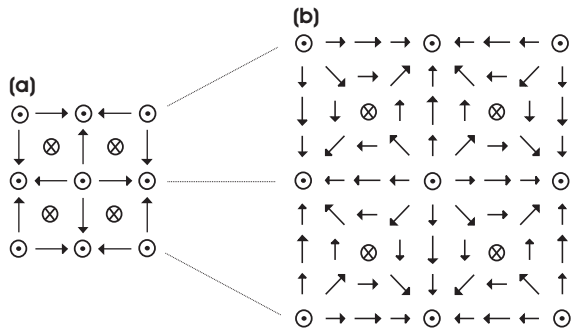}
\caption{ \label{Fig3} Basic cell for the texture of the
magnetization~(a). Every arrow corresponds to the magnetization
direction at a single lattice site. By scaling up once, a more
smoothly varying magnetization texture is obtained, resulting in
smaller solid angles subtended by a single lattice cell~(b).
In~(a) all arrows lie in the plane of the paper, while in~(b) the
magnetization is shown projected onto this plane: smaller arrows
mean a shorter projection.}
\end{figure}
Before discussing the main results of the paper, a few words
should be said about the model we have chosen for the magnetic
texture in our numerical calculations. We started from a $4 \times
4$ cell like depicted in Fig.~\ref{Fig3}a. Only the direction of
the magnetization changes from site to site like depicted in the
figure, while its magnitude stays constant. However, when mapping
this texture to a flux model (like described in
section~\ref{topologicalHall}), one finds that the $4 \times 4$
cell comprises a total flux of $ 4 \, \Phi_0$, corresponding to an
effective flux per lattice cell of $\Phi = 1/4 \Phi_0$, which is
quite large. One could decrease this value by decreasing the solid
angle subtended by the magnetization direction vectors at the
corners of each lattice cell, since the flux per lattice cell is
proportional to it [see Eq.~(\ref{topflux})]. In order to do so
one can scale up the cell of Fig.~\ref{Fig3}a by introducing extra
lattice sites in between the original sites, and by interpolating
the magnetization direction at the new sites between the
directions of the neighboring sites. Scaling up the cell of
Fig.~\ref{Fig3}a once, one then finds a magnetization cell of $8
\times 8$ sites, like depicted in Fig.~\ref{Fig3}b. Of course the
total flux comprised by this cell is still $4 \, \Phi_0$, but this
time distributed over $64$ sites, which corresponds to an
\emph{average} flux per lattice cell of $1/16 \, \Phi_0$. This
magnetizion cell can then in turn again be scaled up; every time
we scale up the cell, it will comprise 4 times the original number
of sites so that the average flux per lattice cell will be
decreased by four. It should be noted that although the effective
flux per lattice cell for the $4 \times 4$ magnetization unit cell
we started from is homogeneously distributed (exactly $1/4 \,
\Phi_0$ per lattice cell), this is not anymore the case for the
cells found by scaling up the first one.

As a first step, we will have a look at a small sections with the
geometry in Fig.~\ref{Fig2}. In the rest of the paper, we will
model these sections by a tight-binding lattice comprising $65
\times 65$ sites. The leads connected to them have a width of $30$
sites. The magnetization in the leads connected to the sample is
chosen to point out of the plane of the paper on every site. When
translating this to the flux model, it means that there is zero
magnetic flux per lattice cell in the leads. In the device itself,
a magnetization cell of $16 \times 16$ sites is used, found from
scaling up the original cell in Fig.~\ref{Fig3}a two times. We can
fit $16$ such magnetization cells in one of our $65 \times 65$
samples.
\begin{figure}
\includegraphics[width = 7.5cm]{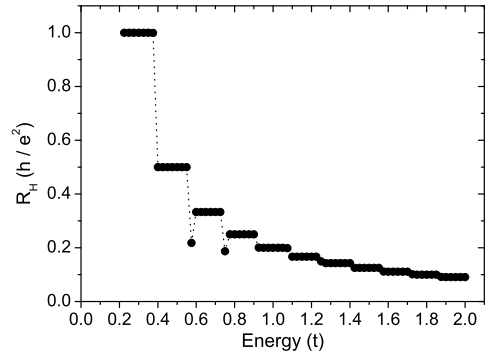}
\caption{ \label{Fig4} Hall resistance as a function of energy for
a single section with magnetization texture obtained by scaling up
the cell in Fig.~\ref{Fig3}a twice.}
\end{figure}

In Fig.~\ref{Fig4}, we have calculated the Hall resistance $R_H$
as defined in Eq.~(\ref{RH}) for a single structure like depicted
in Fig.~\ref{Fig2}, and for energies of the incoming electrons
ranging from $0$ up to $2 \, t$ above the bottom of the spin up
band. The spin splitting due to the magnetization was chosen as
large as $\Delta = g M = 100 \, t$, making sure that we are in the
adiabatic regime (no disorder is present in the system). The
results clearly show a nonzero Hall resistance $R_H$; in fact, one
can clearly observe the integer quantum Hall effect. This indeed
proves that a Hall effect is present although we do not have any
magnetic flux present, nor any kind of spin-orbit coupling. The
topological Hall effect seen here is purely due to the Berry phase
an electron picks up when moving through a structure with a
magnetization texture with nonzero spin chirality.

As explained in section~\ref{topologicalHall}, one can map the
model with the magnetization texture to a spinless flux model. For
the magnetization texture we have chosen, the effective average
flux per lattice cell will be $\Phi / \Phi_0 = 1/64$. This
corresponds to a cyclotron radius of $10 \, a$ (for the Fermi
energy $E_F = 1 \, t$), which is much smaller than the sample size
so that the integer quantum Hall effect could indeed be expected.
When doing the mapping onto the flux model by calculating the
effective hopping parameters like in Eq.~(\ref{teff}) for our
particular choice of magnetization texture, and then calculating
the Hall resistance $R_H$, we found indeed perfect overlapping
between the resistances of the two models. It can be claimed that
the perfect overlap between the two models does not come as a
surprise because in the quantum Hall regime the Hall resistance is
quantized into exact plateaus. However, calculations were done
even outside the quantum Hall regime, and the overlap between the
two models is always exact within our numerical accuracy.
Furthermore, the main point here is the fact that a Hall effect
can be observed that is due to the electron adiabatically
following a certain magnetization structure, and this is clearly
demonstrated by Fig.~\ref{Fig4}.

Now that the topological Hall effect has been demonstrated
numerically, we can proceed to study the transition from the
non-adiabatic to the adiabatic regime. This will be done as a
function of disorder strength (mean free path), in order to check
which of the two adiabaticity criteria in
section~\ref{adiabaticity} are correct. As disorder is introduced
in the system (within the Anderson model), a phase averaging
procedure like set out in the previous section (and described
fully in the appendix) has to be done in order to get quantitative
information. For every separate value of the mean free path, we
have calculated the transmission coefficients for $500$ small
sections (Fig.~\ref{Fig2}) with different impurity configurations.
Then we have wired together $4900$ sections randomly chosen from
these $500$ in a square $70 \times 70$ array, and calculated the
resistivity of the obtained structure like described in the
appendix (Fig.~\ref{Fig7}d). For the magnetization texture, we
have chosen a $64 \times 64$ cell found by scaling up the cell in
Fig.~\ref{Fig3}a three times. A single section thus comprises a
single cell of the magnetization. The shortest distance over which
the magnetization changes its direction by an angle $\pi$ is given
by $\xi \approx 22 \, a$ for this particular cell. The Fermi
energy was fixed to $E_F = 1 \, t$ above the bottom of the spin up
subband.
\begin{widetext}
\begin{center}
\begin{figure}
\includegraphics[width = 17cm]{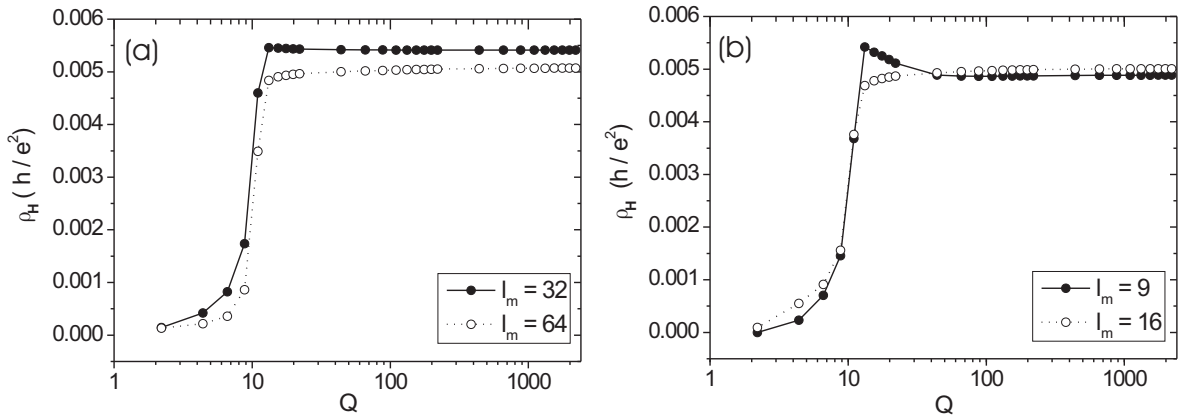}
\caption{ \label{Fig5} Hall resistivity as a function of the
adiabaticity factor $Q$ for different values of the mean free
path.}
\end{figure}
\end{center}
\end{widetext}

Fig.~\ref{Fig5}a shows the Hall resistivity $\rho_H$ defined in
Eq.~(\ref{rhoH}) for mean free paths $l_m = 32 \, a$ and $l_m = 64
\, a$, as a function of the adiabaticity parameter $Q = \omega_s
\xi /v_F = \hbar \Delta \xi / v_F$. For both values of the mean
free path the adiabatic limit is reached simultaneously for values
of $Q \approx 20$, where the Hall resistivity becomes independent
of $Q$. This is in good agreement with the adiabaticity criterium
in Eq.~\ref{critball}. Furthermore, it was checked that the
adiabatic value of the Hall resistivity coincides with the one
calculated by mapping the magnetization model to the flux model.
It should also be noted that $\rho_H$ is independent of the mean
free path ($\rho_H \approx 5 \times 10^{-3} \, \frac{h}{e^2}$)
which is essentially what one might expect from a simple Drude
model. Another feature on the figure is that $\rho_H$ does not
increase linearly for small $Q$; instead it stays very small up to
some point where it changes abruptly. This point coincides with a
value for the spin splitting $\Delta = E_F$. Thus adiabaticity is
reached on a rather short scale as soon as the Fermi energy lies
below the spin down subband.

\begin{figure}
\includegraphics[width = 8cm]{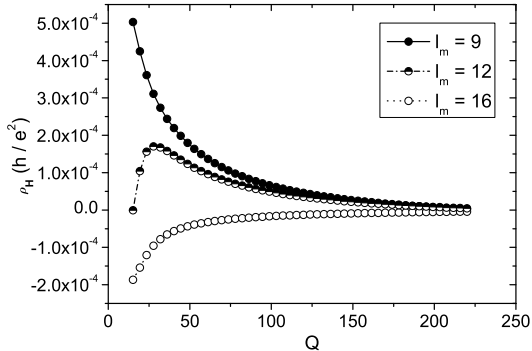}
\caption{ \label{Fig6} Hall resistivity as a function of the
adiabaticity factor $Q$ in the intermediate regime. Decreasing the
mean free path leads to a slower convergence to the adiabatic
value. All plots are shifted so as to converge to $0$.}
\end{figure}

Figure~\ref{Fig5}b shows the same plot, but now for mean free
paths in the diffusive regime ($l_m < \xi$), namely $l_m = 9 \, a$
and $l_m = 16 \, a$. Again, the Hall resistivity increases
abruptly around $Q \approx 20$ to its adiabatic value. The
resistivity $\rho_H$ obtains again the same value as in
Fig.~\ref{Fig5}a around $\rho_H \approx 5 \times 10^{-3} \, h/e^2$
and is thus clearly independent of the mean free path. Although
$\rho_H$ changes abruptly, it can be seen that the real adiabatic
value is reached slower for the lower of the two mean free paths
($l_m = 9 \, a$) for which $\rho_H$ first overshoots its adiabatic
value, and then slowly converges to it. This difference is made
more visible in Fig.~\ref{Fig6}, where we plotted the difference
between the Hall resistivity $\rho_H$ and the adiabatic value it
reaches (so that all curves converge to 0), for mean free paths of
$l_m = 9\, a,12\, a$ and $16\, a$. Again, it is clear that
increasing the amount of disorder leads to a slower convergence to
the adiabatic regime. As such, we confirm that the pessimistic
adiabaticity criterium, which states that adiabaticity is reached
more slowly upon decreasing the ratio $l_m / \xi$ [see
Eq.~(\ref{critpess})], is the correct one. Although for our
limited range of allowed parameters (we should have $l_m >
\lambda_F$ so that localization effects do not start playing a
role) both adiabaticity criteria do not differ very much
quantitatively, $Q \gg 0.45$ versus $Q \gg 2$ for $l_m = 9 \, a$,
the optimistic criterium would predict the opposite behavior.

\section{Conclusions}
In this paper, we have shown numerical calculations confirming the
existence of a fully topological Hall effect, that is due to the
Berry phase an electron picks up when moving adiabatically in a
non-coplanar magnetization texture. In the adiabatic regime, the
governing Hamiltonian can be mapped onto a model of spinless
electrons moving in a magnetic flux; both models indeed give the
same results for the Hall resistance/resistivity. A closer look at
the transition point between non-adiabatic and adiabatic regime
revealed a rather abrupt transition upon increasing the exchange
splitting. The transition takes place around the point where the
spin-down subband becomes depopulated. Furthermore, we were able
to find confirmation for the pessimistic adiabaticity
criterium~\cite{Stern} by looking at the transition point for
different mean free paths in the strongly diffusive regime. In
this regime, a special method for phase averaging should be
introduced for getting rid of conductance fluctuations, which
enables us to describe the transport properties of a large system
($L > L_\phi$) in terms of a Hall resistivity.

\appendix

\section{Phase averaging procedure}
The main idea behind our averaging procedure is that a macroscopic
system can be built up from smaller phase coherent subsections
that are attached in an incoherent way. The small sections we
would like to start from were depicted in Fig.~\ref{Fig2}.
However, there is a subtle point that should be considered first:
it is known that at every interface between a lead and the
electron reservoir feeding this lead, a so-called contact
resistance is present (see e.g.~\cite{Imry} and references cited
there). Essentially such contact resistance results because on one
side (the reservoir) current is carried by an infinite number of
modes, while on the other side (the mesoscopic lead) there are
only a finite number of modes transporting current. When building
up a system of macroscopic size from the smaller sections, such
contact resistance effects in the current-voltage relations of a
single section are unwanted since they are characteristic of the
leads, and not the sample itself.

\begin{figure}
\includegraphics[width = 8 cm]{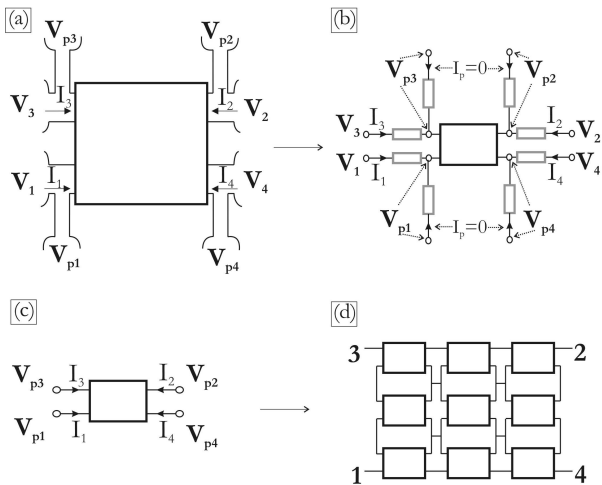}
\caption{ \label{Fig7} Scheme of the phase averaging procedure.
Four extra voltage probes are attached to the original
four-terminal structure~(a). At every reservoir/lead interface a
contact resistance is present; these are depicted in grey in~(b).
Since the voltage probes do not draw current, they will measure
voltage drops over the sample, not including the contact
resistances. Every section can now be considered as an effective
four-terminal box characterized by a set of equations between
currents through the original four leads, and the voltages
measured by the extra voltage probes~(c). Such boxes are wired up
to form a macroscopic system, and its resistivities can be found
by the van der Pauw technique~(d).}
\end{figure}

In order to get rid of contact resistance effects, we will make an
eight-terminal structure by attaching four extra voltage probes as
in Fig.~\ref{Fig7}a. Since these extra voltage probes do not draw
any current, there will be no voltage drop over their contact
resistances, and they will measure the voltage at the point where
they are attached. Placing them like in Fig.~\ref{Fig7}a, one is
able to measure the voltage drops over the sample, excluding the
voltage drops over the contact resistances of the original four
current carrying leads (Fig.~\ref{Fig7}b). Relating now the
currents through the original leads to the voltages measured with
the voltage probes, one gets rid of the contact resistance
effects.

Formally, this can be done with the Landauer-B\"{u}ttiker
formalism as follows. First we write down a set of linear
equations relating the currents and voltages at all eight leads:
\begin{equation}
\left[
\begin{array}{c}
\mathbf{I} \\ \mathbf{I_p}
\end{array}
\right] = \left[
\begin{array}{cc}
A & B \\
C & D
\end{array}
\right] \left[
\begin{array}{c}
\mathbf{V} \\ \mathbf{V_p}
\end{array}
\right],
\end{equation}
where $\mathbf{I} = (I_1,\ldots,I_4)$ is a vector containing the
currents through the original four terminals, and $\mathbf{I_p} =
(I_{p1},\ldots,I_{p4})$ are the currents through the voltage
probes; the same notation convention is used for the voltages on
the leads. The $4 \times 4$ matrices $A,B,C$ and $D$ consist of
transmission coefficients between all eight leads, and are found
directly from the Landauer-B\"{u}ttiker equations.

Since the voltage probes do not draw current, we find
\begin{equation}
\mathbf{I_p} = 0 = C \mathbf{V} + D \mathbf{V_p},
\end{equation}
which can be used to express the currents through the four
current-carrying leads as a function of the voltage on the
attached voltage probes:
\begin{equation} \label{efffourterminal}
\mathbf{I} = \big[ B - A C^{-1} D \big] \mathbf{V_p}.
\end{equation}
Doing so, we have found current-voltage relations for the original
\textit{four-terminal} structure, but since we relate currents
through the original leads to voltages at the voltage probes, we
have gotten rid of contact resistances.

Subsequently, a large number of such structures (with different
impurity configurations) are wired together as shown in
Fig.~\ref{Fig7}d. Every single section is treated as a classical
four-terminal circuit obeying current-voltage relationships of the
form~(\ref{efffourterminal}) [see also Fig.~\ref{Fig7}c]. Doing
so, we introduce an effective phase breaking event at every
connection point between neighboring sections: only current and
voltage information is kept at these points, and any phase
information of electrons flowing out of the section is lost. In
other words, an artificial phase coherence length $L_\phi$
corresponding to the length of a single small section is
introduced in the system.

By applying the correct current conservation laws at the
intersection points in Fig.~\ref{Fig7}d, one can derive a
relationship between the currents and voltages on the four
terminals at the corners of the connected structure (labeled 1 to
4 in the Fig.~\ref{Fig7}d). When enough sections are attached,
this relationship will be independent of the impurity
configurations in the isolated sections. The properties of the
system can then be expressed in terms of resistivities by making
use of the van der Pauw technique~\cite{vanderPauw}. First, the
four-terminal resistances $R_{12,34}, R_{34,12}, R_{14,23}$ and
$R_{42,31}$ between the corners of the large structure are
calculated. With these, the Hall resistivity is defined
by~\cite{vanderPauw, Janecek}
\begin{equation} \label{rhoH}
\rho_H = R_H = \frac{1}{2} ( R_{12,34} - R_{34,12} )
\end{equation}
while the longitudinal resistivity $\rho_L$ can be found from
solving the equation~\cite{vanderPauw}
\begin{equation} \label{rhoL}
\exp \left( - \pi R_{14,23} / \rho_L \right) + \exp \left( - \pi
R_{42,31} / \rho_L \right) = 1
\end{equation}

A small comment should be made here. Because of the attachment
procedure described above, one cannot expect the resistivities
calculated in Eqs.~(\ref{rhoH}) and~(\ref{rhoL}) to correspond
exactly to the real resistivities of a "bulk" system of the same
size and with the same mean free path; in particular, the
resistivities as calculated above depend on the width of the leads
attached to the small sections that make up the large system, and
also on the scheme of wiring the sections together. In theory one
could take into account these effects (see e.g.
Ref.~\onlinecite{vanderPauw}), and one would find that the real
``bulk'' resistivity and the resistivity calculated above are
equal up to a factor that is purely geometrical. Calculating this
factor explicitly is practically quite difficult; we did not
proceed in this direction since the factor is of a purely
geometric origin and has no physical implications.


\begin{thebibliography}{99}
\bibitem{Karplus}
R.~Karplus and J.~M.~Luttinger, Phys.~Rev. \textbf{95}, 1154
(1954) \\
J.~M.~Luttinger, Phys.~Rev. \textbf{112}, 739 (1958).
\bibitem{Smit}
J.~Smit, Physica (Amsterdam) \textbf{24}, 39 (1958).
\bibitem{Berger}
L.~Berger, Phys.~Rev.~B \textbf{2}, 4559 (1970); \emph{ibid.}
\textbf{5}, 1862 (1972).
\bibitem{Hirsch}
J.~E.~Hirsch, Phys. Rev. Lett. \textbf{83}, 1834 (1999).
\bibitem{Taguchi}
Y.~Taguchi, Y.~Oohara, H.~Yoshizawa, N.~Nagaosa, and Y.~Tokura,
Science \textbf{291}, 2573 (2001).\\
Y.~Taguchi, and Y.~Tokura, Europhys.~Lett. \textbf{54}, 401
(2001).
\bibitem{Berry}
M.~V.~Berry, Proc.~R.~Soc.~London A \textbf{392}, 45 (1984).
\bibitem{Bruno}
P.~Bruno, V.~K.~Dugaev, and M.~Taillefumier, Phys.~Rev.~Lett.
\textbf{93}, 096806 (2004).
\bibitem{Ohgushi}
K.~Ohgushi, S.~Murakami, and N.~Nagaosa, Phys.~Rev.~B \textbf{62},
R6065 (2000).
\bibitem{Nagaosa}
S.~Onoda, and N.~Nagaosa, Phys.~Rev.~Lett. \textbf{90}, 196602
(2003).
\bibitem{Tatara}
G.~Tatara, and H.~Kawamura, J.~Phys.~Soc.~Jpn. \textbf{71}, 2613
(2002). \\
M.~Onoda, G.~Tatara, and N.~Nagaosa, J.~Phys.~Soc.~Jpn.
\textbf{73}, 2624 (2004).
\bibitem{Stern}
A.~Stern, Phys.~Rev.~Lett. \textbf{68}, 1022 (1992).
\bibitem{Loss}
D.~Loss, H.~Sch\"{o}ller, and P.~M.~Goldbart, Phys.~Rev.~B
\textbf{48}, 15218 (1993). \\
D.~Loss, H.~Sch\"{o}ller, and P.~M.~Goldbart, Phys.~Rev.~B
\textbf{59}, 13328 (1999).
\bibitem{Anderson}
P.~W.~Anderson, and H.~Hasegawa, Phys.~Rev. \textbf{100}, 675
(1955).
\bibitem{Peierls}
R.~E.~Peierls, Z. Phys. \textbf{80}, 763 (1933).
\bibitem{Popp}
M.~Popp, D.~Frustaglia, and K.~Richter, Phys.~Rev.~B \textbf{68},
R041303 (2003).
\bibitem{vanLangen}
S.~A.~van~Langen, H.~P.~A.~Knops, J.~C.~J.~Paasschens, and
C.~W.~J.~Beenakker, Phys.~Rev.~B \textbf{59}, 2102 (1999).
\bibitem{Buettiker}
M.~B\"{u}ttiker, Phys.~Rev.~Lett \textbf{57}, 1761 (1986).
\bibitem{Buettiker2}
M.~B\"{u}ttiker, Phys.~Rev.~B \textbf{32}, 1846 (1985).
\bibitem{vanderPauw}
L.~J.~van~der~Pauw, Philips~Res.~Repts. \textbf{13}, 1 (1958).
\bibitem{Janecek}
I.~Jane{\v{c}}ek, and P.~Va{\v{s}}ek, Physica C \textbf{402}, 199
(2004).
\bibitem{Imry}
Y.~Imry, \textit{Introduction to Mesoscopic Physics}, (Oxford
Univ. Press, New York, 1997).
\end{thebibliography}
\end{document}